\documentstyle[multicol,pre,aps,epsf,psfig]{revtex}
\begin{document}

\newcommand{\be}{\begin{equation}}
\newcommand{\ba}{\begin{eqnarray}}
\newcommand{\ee}{\end{equation}}
\newcommand{\ea}{\end{eqnarray}}
\newcommand{\ct}{\cite}
\newcommand{\bi}{\bibitem}
\title
{Dynamics of linear polymers in random media}

\author
{Bikas K. Chakrabarti$^{1,2}$, Amit K. Chattopadhyay$^{1}$ and Amit Dutta$^{1}$
}
\address
{
$^1$ Max-Planck-Institut f\"ur  Physik  komplexer Systeme, N\"othnitzer
Strasse 38, 01187 Dresden, Germany \\
$^2$ Saha Institute of Nuclear Physics, 1/AF Bidhannagar, Kolkata-70064,
India
}
\maketitle
\begin{abstract}
We study   phenomenological scaling theories of the polymer dynamics in
random media,  employing the existing
 scaling theories of polymer chains and the percolation statistics.
We investigate both the Rouse and the Zimm model for Brownian dynamics
and estimate the diffusion constant of the center-of-mass of the chain 
in such disordered media. For internal dynamics of the chain, we estimate
the dynamic exponents. We propose similar scaling theory for the reptation
dynamics of the chain in the framework of Flory theory for the disordered
medium. The modifications in the case of correlated disordered
are also discussed.
\end{abstract}

\begin{multicols}{2}
\noindent The conformational statistics of linear polymers  and their dynamical 
properties in porous or disordered media are being investigated 
extensively far a long time [1-14]. 
The universal scaling properties of a linear polymer chain in a good
solvent, which is ideally  modelled in terms of the self-avoiding walk (SAW)
on a regular lattice, are now well established \ct{degennes79,doi86}. However the situation is far from clear in the domain of a polymer moving in a disordered medium.
Knowledge about the  
effects of structural disorder of the medium (or the lattice) 
on the polymer (or
the SAW statistics) and on its dynamical properties are important both for
understanding of the general influence of disorder on the critical
behaviour and also for applications.

The effect of structural disorder of the lattice on the SAW statistics,
in particular when disorder itself is not critical, is still under
investigation. The initial indication \ct{bkc81}
 of the instability of the pure
SAW fixed point in the effective $n$-vector model, was latter shown
to be inappropriate \ct{harris83} in the SAW ($n\to 0$) limit.
However, it seems to be established 
by now \ct{vand92,grassberger93,redner93} 
that the  proper considerations of the
quenched structural disorder of the underlying lattice (fixing one or
both ends of the SAW on the dilute lattice \ct{grassberger93}) indeed induces a modified critical
behaviour for the static conformational statistics of SAWs on disordered
lattices. 
The polymer size exponent $\nu$ is defined through the relation
$R_G \sim N^{\nu}$ connecting the radius of gyration (end-to-end distance
of the SAW) $R_G$ of the polymer chain
 to the molecular weight 
(SAW chain length ) $N$.
The disordered media are usually modelled
as randomly diluted lattices with lattice site or bond concentration
$p$. It is found  that $\nu_d (\equiv \nu (p <1 )) $ is greater than the pure
SAW size exponent $\nu ( \equiv \nu (p=1))$. 
Of course, at
the percolation threshold $p = p_c$, the underlying lattice becomes a fractal
with effective dimensionalities less than the lattice dimension $d$. The
SAW size exponent $\nu_{p_c} (\equiv \nu (p=p_c))$ is therefore clearly
different \ct{lee88,harris89} and in fact is larger than $\nu$ and $\nu_d$.
The multi-fractal effects of the percolation cluster on the SAW conformational
statistics at $p =p_c$ has also been investigated recently \ct{havlin00}.
As mentioned already, all these investigations  attempt to describe
 the static properties
of  polymers  on random lattices.

Compared to the above investigations on the static properties of the
polymer configurations on disordered lattices, not much is known about
the effects of quenched lattice disorder on the dynamics of such
linear polymers \ct{baum96}. Recently, of course, the scaling theory 
and simulations for the reptation dynamics of polymer chains in weakly
disordered lattices ($p \simeq 1)$ has been investigated 
\ct{redner93,cule98}.  The dynamics of un-binding of polymers in a random
media has also been studied \ct{smb97} and the constrained polymer dynamics
has also been investigated using a variational method \ct{blumen01}.
In this report, we propose  scaling theories of polymer
dynamics in such porous or disordered medium using the scaling theory for
polymers \ct{degennes79,doi86} and the percolation statistics 
\ct{stauffer92}. 

We first consider the Rouse model \ct{degennes79} where the interactions
are local. The Brownian dynamics
of the centre of mass $\vec R_{G}$ can be  written as \ct{doi86}

\ba
&~&\frac {d R_G}{dt} \propto f(t); \\ 
&~&<f(t)>=0, <f(t)f(t')> =2 k_B T \delta (t - t'), \nonumber
\ea
where $f(t)$ denotes the temperature ($T$) dependent the random stochastic
noise. One thus  gets
\be
(\Delta R_G)^2 \sim \frac {k_B T } { N} t,
\ee
for the average fluctuation in $R_G$. This gives the diffusion constant
$D \equiv  \Delta R_G^2/t = k_B T/N$.
In the node-link-blob model \ct{stauffer92} describing the percolation
cluster,  the occupied sites or bonds form a super-lattice with
lattice constant $\xi_p \sim (p- p_c)^ {-\nu_p}$, where $\xi_p$ is the
percolation correlation length. In this model,  however another length-scale, 
namely ``chemical length" (along the lattice edge) scaling as
$l_p \sim (p - p_c)^{-\zeta_p}$, appears due to the tortocity of the percolating
paths. Moreover, there exists
random blobs (multiply connected
regions) and dangling ends. Assuming the possibility of Brownian diffusion
on such super-lattices,  we find that the diffusion constant $D$ of a
Gaussian chain can be written as (cf. \ct{degennes79,doi86}).

\ba
D &\sim&   \frac { \frac {(\Delta R_G)^2} 
{ \xi_p^2}}{t} \sim (p - p_c)^{2 \nu_p}
\left(\frac {(\Delta R_{G})^2}{t} \right) \nonumber \\ 
&\sim& (p - p_c)^{\alpha_R}\left(\frac {k_B T}{N}\right)
\ea
where  
\be
\alpha_R = 2\nu_p. 
\ee
Here the radius of gyration $R_G$  
is scaled 
appropriately 
on the percolation cluster 
(super-lattice at $ p > p_c $) by $ \xi_p $  
:$R_G \to R_G/\xi_p$.
It may be noted here that the chain-length $N$ appears here in the
above diffusion equation as mass and thus remains unchanged and
should not be scaled with the chemical length $l_p$ (see however
the later discussion on the reptation dynamics). 
Clearly, the exponent $ \alpha_R $ is always positive 
and therefore $ D $ decreases to zero here as 
$ p \rightarrow p_c $. 

If one considers the long-ranged (hydrodynamic) interactions on top of
the Rouse-like interaction, as in the Zimm-model, 
one gets (cf. \ct{degennes79})

\begin{eqnarray}
D &\sim& \frac { \frac {(\Delta R_G)^2} { \xi_p^2}}{t}
 \sim (p - p_c)^{2 \nu_p}
\left(\frac {(\Delta R_{G})^2}{t}\right) \nonumber\\ 
&\sim& (p-p_c)^{2\nu_p} \left(\frac{k_B T}{N}\right) 
\int dr\:r^{(d-1)} g(r)\:\mu(r) \nonumber \\
 &\sim& (p-p_c)^{2\nu_p}\left(\frac{k_B T}{(R_G/\xi_p)} \right) 
\sim (p-p_c)^{\alpha_Z} 
\left(\frac{k_B T}{N^{\nu_d}} \right)
,
\end{eqnarray}
using similar scaling of the appropriate variables. Here 
$ g(r) \sim [N/{R_G}^d]\:g({r}/{R_G}) $ 
denotes the pair correlation function with $R_G$ scaled by
$\xi_p$  and $ \mu(r) 
\sim r^{-1} $ denotes the mobility function. 
We therefore get

\be
\alpha_Z = \nu_p.
\ee
Here also the 
centre of mass diffusion constant $ D $ decreases to zero
 following the above power law, as $ p \rightarrow p_c $. 

For the internal dynamics or local excitations of the polymer chain,
 one can estimate the dynamic exponent $ z $
  following the same scaling argument \ct{doi86} for the pure case. Writing
   the dynamical structure factor as 

\begin{eqnarray}
g(k,t) &=& {N} g\left(k (\frac {R_G}{\xi_p}),t 
\frac{D}{{(R_G/\xi_p)}^2} \right) \nonumber \\
&\sim& Nk\left(\frac{R_G}{\xi_p} \right)^{-\frac{1}{\nu_d}} 
\tilde g\left(tD (\frac{R_G}{\xi_p})^{-2} [k(\frac{R_G}{\xi_p})]^{z}\right) \nonumber\\
&\sim & \tilde g(tk^z D {R_G}^{z-2} \xi_p^{2-z}) \nonumber\\
&\sim&  \tilde g((p-p_c)^{\gamma}k^zt).
\end{eqnarray}
Since the local excitations (for $ k R_G \gg 1 $) are independent of $ N $, 
assuming $ D \sim (p-p_c)^{\alpha} N^{-\beta}$(with $\alpha= 2 \nu_p, \beta=1$ 
 for the
 Rouse model (Eq.~(3)) and 
$\alpha = \nu_p$, $\beta = \nu_d$ (Eq.~(6)) for the Zimm model, one gets 

\be
z=2+\frac{\beta}{\nu_d}, ~~{\rm and}~~ \gamma = \alpha + \nu_p(z-2).
\ee
Hence $ z=3 $ for all dimensions in the Zimm model, 
while $ z $ changes to $ 2 + 1/\nu_d $ for the disordered 
case in the Rouse model and its magnitude becomes equal to 4 for $ d \geq 6 $.
 
We have assumed in the above discussion, the possibilities of the Rouse or
Zimm dynamics of the polymer chain in good solvents within the pores of the
random media. This obviously necessitates the correlated percolation picture of 
the porous media, and hence the appropriate modifications \ct{ferber01} of the
polymer statistics in such medium (see below). However, very near the
percolation threshold, the polymer dynamics can perhaps be only reptation
type \ct{degennes79} along the constrained pore tubes. Following the earlier
investigations \ct{redner93,cule98}, the diffusion constant $D$ for the
polymer, making reptation dynamics in the disordered media, can be written as 

\begin{eqnarray}
D &\sim & \left (\frac {(\frac {R_G}{\xi_p})^2}{\tau} \right) 
\exp \left(- \frac {\Delta F(R_G)}
{k_B T} \right) \nonumber\\
&\sim & \frac {(p-p_c)^{2\nu_p}}{N^{3-2\nu_d}} 
\exp \left(- \frac {\Delta F(R_G)}{k_B T} \right),
\ea
where $R_G \sim N^{\nu_d}$, $\tau \sim N^{3}$ and $\Delta F(R_G)$
denotes the fluctuation in the polymer free energy due to the structural 
disorder of the medium. A simple Flory type estimate for the free energy
$F$ can be written, following Smailer {\it et~al} \ct{redner93}, as 

\be
F \sim \left(\frac{\frac {R}{\xi_p}}{(\frac {N}{l_p})^{\nu}} 
\right)^{1/(1-\nu)} 
+ \left( \frac {(\frac {N}{l_p})^2}{(\frac {R }{\xi_p})^d} \right)^{1/2},
\ee
where $\nu(= 3/(d+2))$ 
is the SAW size exponent on the pure lattice and the second term,
arising due to the random fluctuation of the density-density (excluded
volume) interaction, gives the configurational fluctuation $\Delta F$. 
The most important point here is the rescaling of the polymer mass
with the chemical length $l_p$. Until the above equation, we considered
scaling of only the fluctuation $\Delta R_G$ by the lattice constant
$\xi_p$ of the  percolating super-lattice, and not of the
polymer mass $N$. In the above expression for the Flory 
free energy,  on the other
hand, one has to scale as well the chain length or mass $N$ by
$l_p$ \ct{barat95}. This is necessary here to cast the free energy
in the disorderd media in the pure lattice form; or in other  words,
as in the free energy, both temperature and mass renormalisatins
are allowed.
The minimisation, with respect to $R$ of the above free-energy gives
the polymer size as 

\ba
R_G &\sim& (p-p_c)^{-\delta} N^{\nu_d}; \nonumber\\
\nu_d &=& \frac 2 {2 + d (1 - \nu)}, ~~~
\delta =  \nu_p -\zeta_p \nu_d. 
\ea

\noindent The above result may also be obtained by scaling the 
relation ($R_G \sim N^{\nu_d}$) connecting the radius of gyration to the
 polymer chain length: $R_G \to R_G/\xi_p$ and $N \to N/\l_p$ \ct{akr}. 
The value of the
exponent $\delta$ is positive \ct{stauffer92} for $d=2$ and $3$. This indicates
that the polymer swells (for fixed $N$) 
 as $p \to p_c$. The possible negative value of
$\delta$ for $d \geq 4$ may indicate localisation  of the polymer in
higher-dimnensions (Gaussian chains) in the 
random media \ct{baum96}.   

The fluctuation $\Delta F$ can therefore be written as

\be
\Delta F \sim (p-p_c)^{-\phi} N^{\chi};~~\phi = \frac {d}{2}\nu_p - \zeta_p,~~
\chi = \frac {\nu_d}{2}(2 - d \nu).
\ee
It may be noted from the above expression  that $\nu_d> \nu$  for
all $d < 4$ and we assume $\nu_d = \nu = 1/2$ for $d \geq 4$.
The values of the exponents $\phi$ and $\chi$ are positive
\ct{stauffer92}  for $d<4$. 
The above Eqs.~ (10) and (12) together describes the scaling behaviour of
the diffusion constant ($D$) of the reptation motion for the polymer chain in
the disordered medium. We find that for $d <4$,
 $D$ decreases exponentially not only
as the chain length $N$ increases but also as $p \to p_c$.
 At $d=4$, $\chi$ vanishes while $\phi$
becomes slightly negative so that the diffusion constant $D$ in (12)
becomes independent
of $N$ and decreses with $p$ approaching $p_c$.
This seems to be related to the possible localisation
of polymers in higher-dimensions as discussed already.

As mentioned already, these dynamical properties of polymers in porous
media demand the consideration of the correlated randomness of the percolating
cluster.	
Very recently, Blavasts'ka {\it et~al} \ct{ferber01} have studied the
static scaling properties of polymers on a $d$-dimensional random
lattice with ``correlated" randomness that has a power-law fall 
of the form  $1/r^a$ for large separation $r$ \ct{halperin82}. Similar
Flory-type estimate for the free energy $F$ in the case of relevant
long-range correlation ($a<d$) will be
given by Eq.~(10) where the spatial dimensionality $d$ in the second
term  will be replaced by the parameter $a$ denoting the range of interaction.
For $a \geq 4$, this  fluctuation term becomes irrelevant and thus we
get the mean field critical behaviour. Using therefore
the expression (11) for the Flory estimate of $\nu_d$ for the
non-trivial cases  $a<4$ ($d \simeq 4$), we get 
\be
\nu_d = \frac 2 { 2 + a \left(\frac {d-1}{d+2} \right)} \simeq \frac 1 {2}
+\frac{ \delta}{16},  
\ee
where $\delta = 4-a$ and   $\nu = 1/2 + (4-d)/12$. Clearly, this size
exponent $\nu_d$, obtained via Flory theory, does not match the renormalisation
group result \ct{ferber01} even in the first-order in $\delta$. This is
is  similar to the failure of the Flory approximation for the
exponent $\nu$ in the first-order in $(4-d)$ even in the case of
pure lattice (with short-range interactions) \ct{degennes79}. It may however
 be noted 
that this result for $\nu_d$ is for the
long-range case $a <d$, where the renormalisation group
study in \ct{ferber01} failed to locate any stable fixed point. For $a>d$,
the above method clearly yield the pure (short-range)
 SAW result. However, the renormalisation
group study suggests the existence of a long-range fixed point in this region
\ct{ferber01}.

In conclusion, we  have studied the dynamical properties of 
 lattice models of a linear polymer on a
percolating lattice model of the random or porous media, using
simple scaling ideas for both  the polymer chain and the 
disordered media.
The  results for the  dynamical exponent $z$
describing the internal dynamics, and the diffusion
coefficient $D$ of the center of mass of the polymer chain, using both
the Rouse and the Zimm model, have been investigated for such disordered
media. We find that while the diffusion coefficient decreases with 
$p - p_c$, following power laws in both the cases, the dynamic exponent
$z$ remains unchanged to its pure value 3 for the Zimm model but gets
modified for the Rouse model. 
 Applications of
the Flory theory for the polymer free-energy and the scaling theory
of the percolation disorder indicate that the  reptation diffusion
coefficient $D$ of the polymer chain decreases exponentially with the
chain length ($N$) and also inversely with the deviation ($
p - p_c$) from the percolation threshold $p_c$: $D \sim \exp [- N^{\chi}
(p -p_c)^{\phi}]$,
 where the exponents $\chi$ and $\phi$ are both positive and related to
the polymer size and the percolation exponents respectively. 
Thus, based on a simple Flory-type theory, we have extended a few dynamical properties of linear polymers near the percolation threshold and have also discussed the case of correlated randomness. 
\section * {acknowledgement}

\noindent  BKC is grateful Peter Fulde for his kind hospitality at MPIPKS, Dresden,
Germany.

\end{multicols}
\end{document}